\documentclass[12pt]{iopart}

\usepackage{graphicx}
\begin{document}

\title{Domain boundary formation  in helical multishell gold nanowire}

\author{T. Hoshi$^{1,2}$, T. Fujiwara$^{3,2}$}

\address{(1) Department of Applied Mathematics and Physics, 
Tottori University, Tottori 680-8550, Japan}
\address{(2) Core Research for Evolutional Science and Technology, 
Japan Science and Technology Agency (CREST-JST), Japan}
\address{(3) Center for Research and Development of Higher Education, 
The University of Tokyo, 
Bunkyo-ku, Tokyo, 113-8656, Japan}

\ead{hoshi@damp.tottori-u.ac.jp}

\begin{abstract}
Helical multishell gold nanowire is studied theoretically
for the formation mechanism of helical domain boundary.
Nanowires with the wire length of more than 10 nm
are relaxed by quantum mechanical molecular dynamics simulation
with tight-binding form Hamiltonian.
In results, non-helical nanowires are transformed 
into helical ones with the formation of atom pair defects at domain boundary,
where the defective atom pair is moved from an inner shell.
Analysis of local electronic structure shows 
a competitive feature of the energy gain of 
reconstruction on wire surface and the energy loss of the defect formation.
A simple energy scaling theory
gives a general discussion of domain boundary formation.
\end{abstract}

\maketitle

\section{Introduction}

Nanometer scale material forms exotic structures and 
it is crucial to understand and control their formation mechanism,
so as to establish the foundation of nano electronics.
Helical multishell gold nanowire ~\cite{KONDO2000, OSHIMA2003-JEM}
is one of  exotic metal nanostructures \cite{AGRAIT2003}
and has characteristic multishell configurations with \lq magic numbers'. 
The helical wires were synthesized by focusing electron beam on thin film. \cite{KONDO1997}
The wire axis is in the [110] direction of the original FCC structure and 
the outermost shell is a folded (111)-type (hexagonal) atomic sheet with helicity.
A single shell helical structure was synthesized later. \cite{OSHIMA2003}
The transport property was studied theoretically \cite{ONO2005}
and experimentally \cite{OSHIMA2006}.
Platinum nanowires were also synthesized 
with the same type of helicity. ~\cite{OSHIMA2002}

The observed multishell structures of the helical gold nanowires 
~\cite{KONDO2000} are 
denoted by the numbers of atoms in each shell and are
classified into 7-1, 11-4, 13-6, 14-7-1, 15-8-1 structures.
For example, a \lq 14-7-1 nanowire'  is a nanowire with three shells and
the outer, middle and inner shells have fourteen, seven and one atom(s) 
in the section view, respectively.  
These numbers are called \lq magic numbers',
since the difference of the numbers between the outermost 
and  next outermost shells is seven, except the cases of the 7-1 structure.
 
Theoretical investigations were carried out for the helical nanowire structures 
\cite{Gulseren1998, tosatti2001, BILAL2003, senger2004, yang2004, 
LIN2005, IGUCHI2007, FUJIWARA2008, AMORIM2008} and,  
among them, 
a recent theory ~\cite{IGUCHI2007, IGUCHI-THESIS} explains 
the observed multishell configuration with \lq magic numbers' systematically.
The theory was proposed as 
a two-stage formation model  
of helical multishell gold nanowires 
and the model was confirmed by 
quantum mechanical molecular dynamics
simulations for gold and copper with tight-binding form Hamiltonians.
The simulation shows the energy relaxation process 
from non-helical structures into helical ones. 
The analysis of electronic structure concluded that 
the origin of the helicity comes 
from the intrinsic nature of 
non-spherical $5d$ electrons and 
a (111)-type (hexagonal) surface structure is energetically favorable 
for $5d$ electrons in sheet structure. 
The above mechanism gives 
a general understanding 
among (i) the appearance of helical nanowire structures of gold and platinum and 
(ii) the fact that 
reconstructed equilibrium surfaces  of FCC 5$d$ metals, gold platinum and iridium, 
prefer (111)-type structures. 
\cite{FEDAK-1967, BINNIG-1983, HO1987-Au110, VANHOVE-1981, binnig1984, abernathy1992, JAHN-1999}
After the theory paper, \cite{IGUCHI2007}
several related simulations were carried out 
for formation of helical gold nanowire within tight-binding form Hamiltonian.
\cite{FUJIWARA2008,AMORIM2008}

The present paper investigates the formation of helical domain boundary on  wire surface,
where the fundamental picture and simulation method  are shared 
with the previous theory paper. \cite{IGUCHI2007}
In general, a defect should be introduced at domain boundary
in the formation process of a helical domain from non-helical one. 
The information of domain boundary is missing in the previous  paper,
since the simulated structures are short isolated wires 
of which wire length is less than 3 nm. 
In the resultant helical nanowires, 
the domain boundaries are located in the wire ends that are terminated artificially. 
In the present paper,
simulations are carried out with longer wires, longer than 10 nm, 
and reveal a possible defect induced mechanism for forming helical domain boundary. 

This paper is organized as follows;
Section ~\ref{SEC-SIMULATION} describes 
the methodology and result of the simulation.
Section ~\ref{SEC-ANALYSIS} focuses on 
the analysis of the results, particularly on 
the energy per atom and 
the local density of states (LDOS). 
In Sec.~\ref{SEC-DISCUSSION},
a simple energy scaling theory 
with respect to the wire length of helical domain
is constructed for a general mechanism of domain boundary formation.
Finally, the summary and future aspects are given in 
Sec.~\ref{SEC-SUMMARY}.

\begin{figure}
\begin{center}
\includegraphics[width=0.99\linewidth]{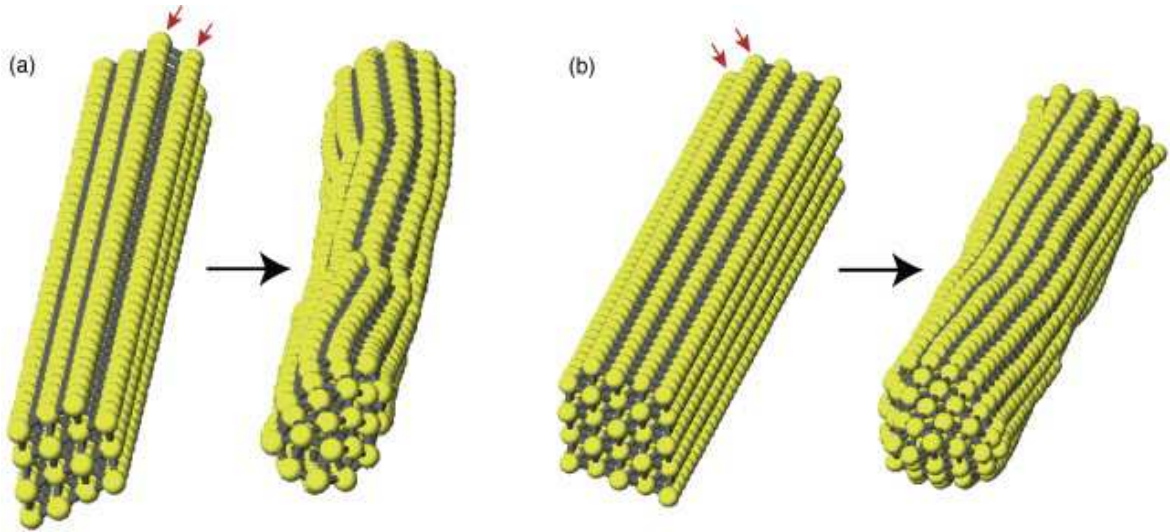}
\end{center}
\caption{
\label{fig-Au608-viewpoint}
Simulated formation process of helical multishell gold nanowires 
with the wire lengths longer than 10 nm. 
The 11-4 structure is depicted in (a)
at the initial structure (left panel) and the 9,000-th iteration step (right panel) and
the 15-8-1 structure is in (b)
at the initial structure (left panel) and the 6,600-th iteration step (right panel).
See text for details. 
}
\end{figure}

\section{Simulation} 
\label{SEC-SIMULATION}

The formation process of the helical wires is realized
by quantum mechanical molecular dynamics simulation 
as a structure relaxation with thermal fluctuation. 
The temperature was set to be $T=$600 K,
lower than the melting temperature (1337K).
The simulation code used here has the name of \lq ELSES'
(=Extra-Large-Scale Electronic Structure calculations).
\cite{ELSES-URL, HOSHI2000, HOSHI2003, 
geshi2004, TAKAYAMA2004, HOSHI2005, TAKAYAMA2006,
 HOSHI2006, IGUCHI2007,  FUJIWARA2008,YAMAMOTO2008, HOSHI-JPCM2009}. 
The simulation was realized  with
a tight-binding form Hamiltonian 
~\cite{MEHL1996, KIRCHHOFF2001, NRL-REVIEW},
which was used 
for gold nanowire.
\cite{dasilva2001, dasilva2004,haftel2006,IGUCHI2007,FUJIWARA2008,AMORIM2008}
The Hamiltonian form contains several parameters and they are determined
to represent electronic structures of bulk solids, 
surfaces, stacking faults and point defects. 
One iteration step corresponds to $\Delta t = 1$ fs. 
Eigen state calculation for electronic structure is carried out at each iteration step.
The boundary condition is imposed by fixing 
the mass center of the wire end layers. \cite{IGUCHI2007}
The computational time with 608 atoms is typically
nine minutes per one iteration step,
when a workstation 
with four dual-core Opteron$^{\rm (TM)}$ processors was used.

We should say that
the present simulation is different from experiment in several points.
For example,
the simulated time is quite short, 7-9 ps, 
owing to the practical limit of computational resource. 
Therefore, 
the simulation result should be understood so that 
it captures an intrinsic energetical mechanism of the real process.

The resultant nanowires 
are shown in Fig.~\ref{fig-Au608-viewpoint}
for the (a) 11-4 and (b) 15-8-1 structures and 
contain multiple helical domains on wire surface
with well-defined domain boundary, 
unlike the smaller samples in the previous paper. \cite{IGUCHI2007} 
The initial structures are parts of ideal FCC structure 
and are the same as those in the previous paper, except their wire length. 
The wire lengh of these nanowires is approximately 12 nm 
or is 40 or 42 unit layers for the 11-4 and 15-8-1 structures, respectively.  
Here a unit layer of a [110] nanowire 
consists of two successive atom layers. 
At the initial structures,
the wire surface consists of (111)-type (hexagonal) and 
(001)-type (square) areas. 
The latter area consists of two atom rows 
that are indicated by the two parallel arrows 
in Fig.~\ref{fig-Au608-viewpoint}.  
Defect structures appear on wire surface as helical domain boundary. 
Hereafter the domain boundary formation 
of the 11-4 structure will be focused on.

\begin{figure}
\begin{center}
\includegraphics[width=0.99\linewidth]{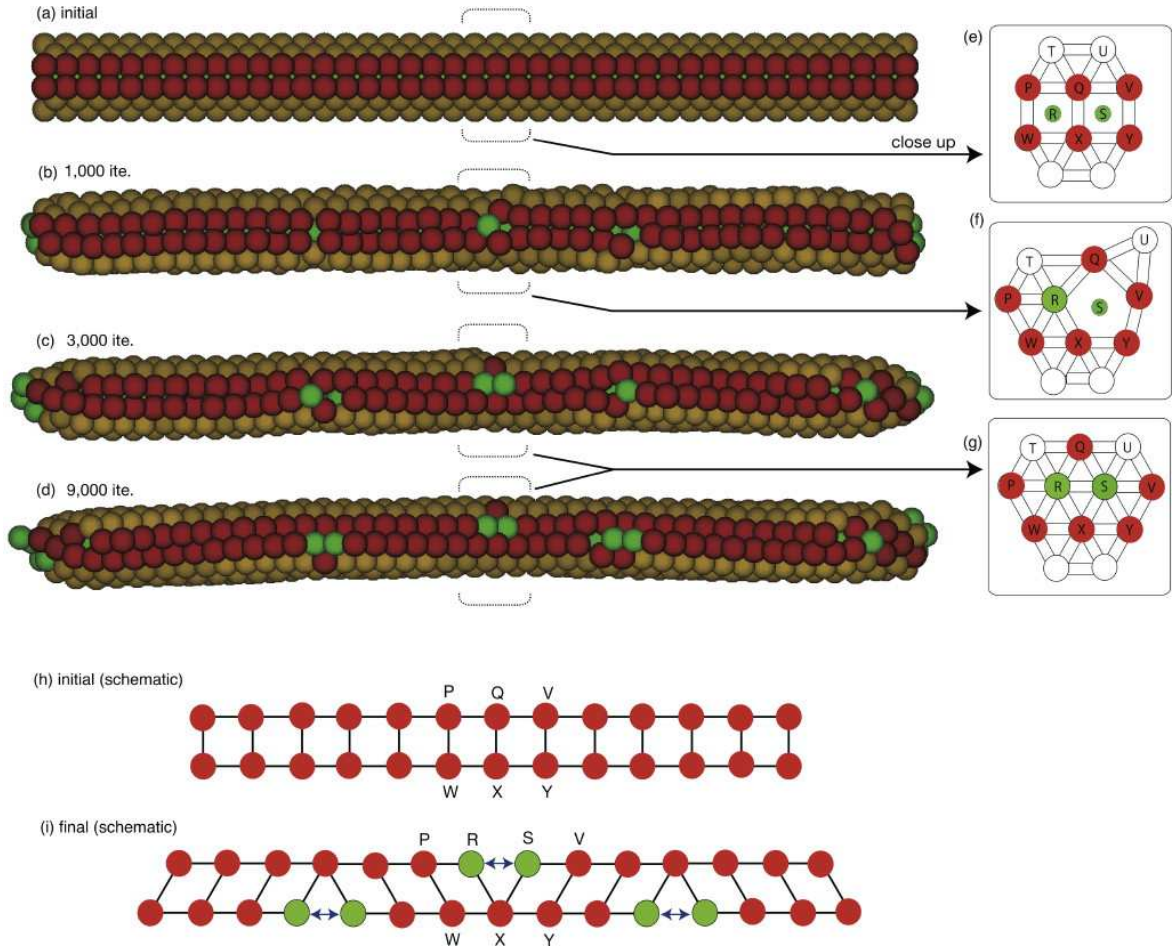}
\end{center}
\caption{
\label{fig-Au608-snapshots}
The transformation process of the 11-4 structure. 
(a) (b) (c) (d) : Snapshots at the initial structure 
and 1,000-th, 3,000-th and 9,000-th iteration steps, respectively.
(e) (f) (g) : Sketched figure as a close up of 
the central area that are indicated by broken lines on the wire surface.
The figures (e) and (f) capture the structures of (a) and (b), respectively and 
the figure (g) captures the structure of (c) or (d). 
The atoms depicted as smaller balls,
$R$ and $S$ in (e) and $S$ in (f), are placed within the inner shell. 
(h) (i) : Schematic figure of formation of atom-pair defects  
in the (h) initial and (i) final structures.
}
\end{figure}

Figure \ref{fig-Au608-snapshots} shows 
several snapshots of sideview for the 11-4 case,
where the atoms are distinguished by color,
so as to clarify the transformation process. 
Red atoms are those initially placed on the (001)-type surface area and
green atoms are those initially placed  in the inner shell region. 
The other atoms are those initially placed on the (111)-type surface area.
The central region indicated by broken lines at Fig.~\ref{fig-Au608-snapshots}(a)-(d)
is sketched in Fig.~\ref{fig-Au608-snapshots}(e) (f) and (g). 
Figure ~\ref{fig-Au608-snapshots}(e) and (f) capture 
the structures of Fig.~\ref{fig-Au608-snapshots}(a) and 
Fig.~\ref{fig-Au608-snapshots}(b), respectively and 
Fig.~\ref{fig-Au608-snapshots} (g) captures the structure of 
Fig.~\ref{fig-Au608-snapshots}(c) or Fig.~\ref{fig-Au608-snapshots}(d). 
At the initial structure, Fig.~\ref{fig-Au608-snapshots}(e), 
the \lq red' atoms,
marked as $P, Q, V, W, X$ and $Y$,
form a square lattice on wire surface and 
the \lq green' atoms, marked as $R$ and $S$, are placed 
in the inner shell. 
At the final snapshot, Fig.~\ref{fig-Au608-snapshots}(g), 
the two \lq green' atoms $R$ and $S$, 
are moved from the inner shell into the wire surface,
which form a helical domain boundary. 
They are inserted between the atoms $P$ and $V$, 
while the atom $Q$ is moved into an upper area between the atoms $T$ and $U$.  
As result,
the surface reconstruction occurs,
from the (001)-type (square) lattice 
into the (111)-type (hexagonal) one,  
and introduces the helical domains with domain boundaries. 

As a remarkable tendency of the entire sample, 
surface defects at domain boundaries 
appear typically as pairs of \lq green' atoms,
as illustrated in Fig.~\ref{fig-Au608-snapshots}(h) and (i). 
The central atom-pair defect,
the pair of the atom $R$ and $S$, appears,
because the shear-like deformation on the (001)-type lattice
occurs in the opposite shear directions 
between the left and right areas of the defect,
as indicated by a blue arrow in Fig.~\ref{fig-Au608-snapshots}(i).

\begin{figure}
\begin{center}
\includegraphics[width=0.99\linewidth]{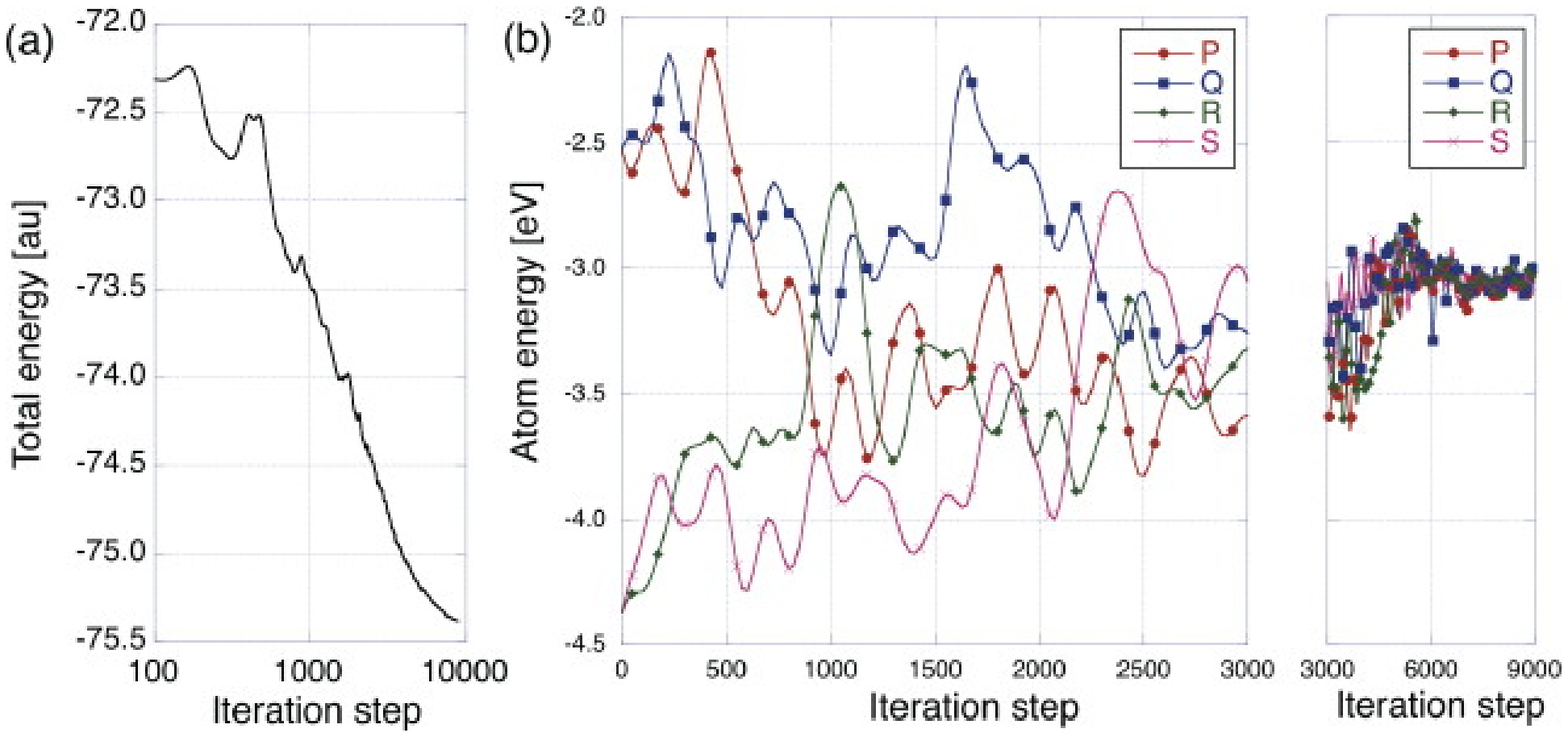}
\end{center}
\caption{
\label{fig-Au608-energy-change}
Energy change of the 11-4 structure during the relaxation process. 
(a) Change of the total energy. 
(b) Change of the atom energy for the atom $P$, $Q$, $R$ and $S$
that are indicated in Fig.~\ref{fig-Au608-snapshots}.
}
\end{figure}

\begin{figure}
\begin{center}
\includegraphics[width=0.7\linewidth]{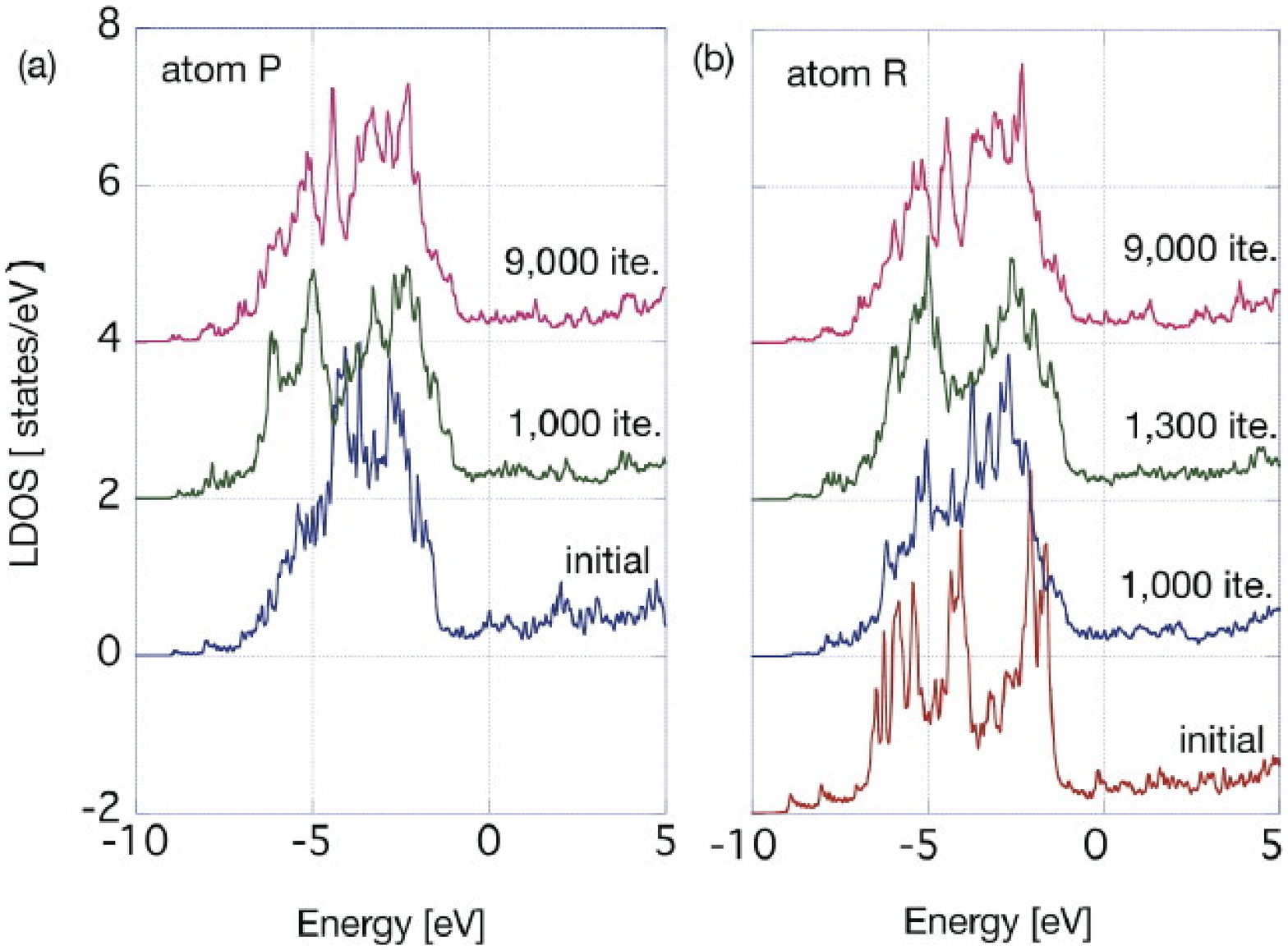}
\end{center}
\caption{
\label{fig-Au608-LDOS}
Local density of states of the atoms $P$ and $R$
at the initial structure and several iteration steps
written in the graphs. 
The origin of the vertical axis is shifted 
among different iteration steps. 
The origin of the horizontal axis is set to the Fermi level.
}
\end{figure}

\section{Analysis of local electronic structure}
\label{SEC-ANALYSIS}

The transformation process from the non-helical wire into
the helical one occurs with energy gain, according to 
the change of the total energy shown in Fig.~\ref{fig-Au608-energy-change}(a). 

The change of the atom energy, energy of each atom,
is plotted in Fig.~\ref{fig-Au608-energy-change}(b)
for the selected atoms $P$, $Q$, $R$ and $S$,
that are indicated in Figs.~\ref{fig-Au608-snapshots} (e), (f) and (g). 
Figure ~\ref{fig-Au608-LDOS}(a)
shows the LDOS of the atom $P$  
at the initial structure 
and the 1,000-th and 9,000-th steps.
Figure ~\ref{fig-Au608-LDOS}(b) shows
the LDOS of the atom $R$
at the initial structure 
and the 1,000-th, 1,300-th and 9,000-th steps.
In Fig.~\ref{fig-Au608-LDOS}, 
the origin of the vertical axis is shifted among the snapshots and 
the origin of the horizontal axis is set to the Fermi level for each snapshot.
Since the values of the Fermi level are different only about or within 0.1 eV 
among the snapshots,
the difference in the Fermi level is negligible in the energy scale of the graphs.
Since the atoms $R$ is placed in the inner shell at the initial structure, 
the LDOS profile, shown in the lowest graph 
of Fig.~\ref{fig-Au608-LDOS}(a), 
is similar to that of bulk FCC gold and 
has characteristic three peaks in the $5d$ band 
at $\varepsilon \approx -2$ eV, -4 eV and -6 eV.

The atom movement from the inner shell into the wire surface 
accompanies the drastic change of the atom energy. 
Figure ~\ref{fig-Au608-energy-change}(b) contains 
temporal peaks nearly at the 1,000-th and 2,400-th iterations 
for the atom $R$ and $S$, respectively,
which corresponds to 
the successive atom movements of the atom $R$ and $S$, 
from the inner shell into the wire surface,
shown in Fig.~\ref{fig-Au608-snapshots}.

When the initial and final structures are compared in the atom energy, 
one can find that the helical transformation 
is motivated by a surface effect. 
At the final iteration step (9,000-th iteration step), 
the four atoms $P$, $Q$, $R$ and $S$
are transformed into members of the folded (111)-type surface with helicity,
as shown in Figs.~\ref{fig-Au608-snapshots} (d) and (g).
Therefore the atom energy of them reaches an almost unique value, 
$E \approx -3.1$ eV in  Fig.~\ref{fig-Au608-energy-change}(b), and 
the final LDOS profile of the atoms $P$ and $R$ are quite similar,
as plotted in the highest graphs of 
Fig.~\ref{fig-Au608-LDOS}(a) and (b).
In Fig.~\ref{fig-Au608-snapshots}, 
the atoms $P$ and $Q$, depicted as \lq red' atoms 
are transformed from the (001)-type surface area into the (111)-type one.
Figure ~\ref{fig-Au608-energy-change}(b) shows that 
the atom energy of the two atoms decreases 
from $ E \approx - 2.5$ eV into $ E \approx - 3.1$ eV. 
The energy gain mechanism is explained 
by the two-stage model in the previous paper and  
the present case corresponds to the case of the atom $B$ in the paper. \cite{IGUCHI2007}
On the other hand, 
the atoms $R$ and $S$, depicted as \lq green' atoms in Fig.~\ref{fig-Au608-snapshots}, 
are moved from the inner shell into the (111)-type surface area.
Figure ~\ref{fig-Au608-energy-change}(b) shows that 
the atom energy of the two atoms increases 
from $ E \approx - 4.3$ eV into $ E \approx - 3.1$ eV.

The above observation in the increase and decrease of the atom energy is confirmed, 
when histograms (not shown) are constructed 
for the atom energy difference  $\Delta E$ between  the initial and final structures.
The histograms are constructed among all the atoms except the 158 atoms 
located within the five layers near the two wire ends,
so as to avoid possible artifacts of the wire ends. 
All the \lq red' atoms, the atoms originally placed on the (001)-type area of the surface,
show the energy decrease of  $\Delta E \approx - 0.2 \sim - 1.2$  eV
and the histogram peak is located at $\Delta E \approx -0.7$ eV. 
The \lq green' atoms, the atoms originally placed in the inner shell,
show the energy difference 
from $\Delta E \approx $ - 0.2 eV to 1.6 eV. 
The large energy increase ($\Delta E \approx  1.2 \sim 1.6$ eV)
comes from the atoms moved into the wire surface, like the atom $R$ and $S$. 
Energy increase also appear among several other \lq green' atoms in the inner shell,
since the resultant helical wire has defective regions in the inner shell,
particularly regions near the moved atoms.  
The histogram of the rest atoms, the atoms originally placed on the (111)-type area of the surface,
shows a wide range of values $\Delta E \approx - 1.2 \sim  1.0$  eV
and their average is an energy decrease ($\Delta E \approx - 0.16$ eV). 
We should remember that these \lq rest' atoms 
are different in their situations in the initial structure.
See the cases of the atom $A$ and $C$ in the previous paper. \cite{IGUCHI2007}

It should be noted that, 
an experimental paper of [110] gold nanowire \cite{KONDO1997},
earlier than the report of helical structure\cite{KONDO2000},
suggests a surface stabilization mechanism 
in which (001)-type area on the wire surface
reconstruct into (111)-type ones as on equilibrium surface.
The suggested stabilization mechanism 
is consistent to the previous theory \cite{IGUCHI2007} and the present analysis. 

The analysis of the intermediate structures reveals 
a competitive feature of temporal energy gain and loss among atoms. 
In Fig.~\ref{fig-Au608-energy-change}(b), 
for example, the atom $R$ shows energy loss 
during the 800-th and 1,000-th iteration step
($E \approx - 3.7$ eV $\Rightarrow -2.7$ eV),
while the atoms $P$and $Q$ show energy gain during the same period
($E \approx - 3.1$ eV $\Rightarrow -3.7$ eV and 
$E \approx - 2.8$ eV $\Rightarrow -3.4$ eV
for $P$ and $Q$ respectively).
The temporal energy gain and loss are almost the same amplitude.
During the period, the atom $R$ is moved from the inner shell to the wire surface 
(See Fig.~\ref{fig-Au608-snapshots} (b)).
 In conclusion, 
the atom $R$ is moved with energy loss, from the inner shell into the wire surface,
because of the energy gain mechanism of the atom $P$ and $R$. 
The energy gain is found also in the LDOS profile of the atom $P$,
shown in Fig.~\ref{fig-Au608-LDOS}(a). 
When the LDOS profile is compared between 
those at the initial structure and at the 1,000-th step,
the weighted center of the $5d$ band is shifted downwards.
Moreover a gap-like structure appears,
at $ \varepsilon \approx -4.5$ eV,
in the LDOS profile of the 1,000-th step,
which indicates the formation of bonding and antibonding states. 
In short, an unstable surface atom is changed 
into a ( relatively ) stable one,
owing to the energy gain of the $5d$ electrons. \cite{IGUCHI2007}
The atom $R$ also shows a temporal energy gain
after the movement into the wire surface,
during the period between 1,000-th and 1,300-th iterations
($E \approx - 2.7$ eV $\Rightarrow -3.8$ eV). 
The change of the LDOS profile during the period,
shown in Fig.~\ref{fig-Au608-LDOS}(b) is similar to 
the change of the atom $P$ discussed above,
which indicates the same stabilization mechanism with the 5$d$ electrons.

\section{Discussion}
\label{SEC-DISCUSSION}

The above analysis indicates that 
the helical transformation is motivated by the energy gain
in the surface reconstruction 
from the (001)-type area the into (111)-type one 
and sacrifices the energy loss 
for forming point defects in the domain boundaries on the wire surface.  

A simple theory is constructed by energy scaling 
with respect to the wire length of a helical domain $L_{\rm dom}$.
The main energy gain appears among \lq red' atoms 
that are distributed on line 
and the energy gain is scaled as $O(L_{\rm dom})$, 
whereas the energy loss in forming the atom-pair defect 
at domain boundary is scaled as $O(1)$. 
From the analysis of the local energy of Fig.~\ref{fig-Au608-energy-change}(b), 
the energy gain for each reconstructed \lq red' atom 
is $E_{\rm gain} \approx 0.6$ eV 
(See the cases of the atom $P$ or $Q$),
while the energy loss of a \lq green' atom of the defect 
is  $E_{\rm loss} \approx 1.2$ eV
(See the cases of the atom $R$ or $S$).
The energy gain of helical domain  is estimated to be $2 E_{\rm gain}$ per unit layer,
since one unit layer of the wire contains two \lq red' atoms.
The energy loss at domain boundary, on the other hand,
is estimated to be $2 E_{\rm loss}$,
since the defect appears typically as an atom pair. 
The above estimation concludes that  
a helical domain should appear,
if the wire length of the domain is enough long to satisfy
$L_{\rm dom} > L_{\rm dom}^{\rm (c)} 
\equiv (2 E_{\rm loss})/(2 E_{\rm gain}) = 2$ unit layers,
which is consistent to the resultant domain structure of  
Fig.~\ref{fig-Au608-snapshots}.

We emphasis that the above scaling theory can be constructed,
because the simulated wire is much longer than
the critical length $L_{\rm dom}^{\rm (c)}$,
unlike in the previous paper,
in which the length is less than ten layers. \cite{IGUCHI2007}
Although the atom movement from the inner shell into the wire surface 
is also found in a shorter wire of the 12-6-1 structure, \cite{IGUCHI2007}
the movement occurs within one atom line and the energy loss 
cannot be scaled as $O(1)$.

The present simulation gives two conclusive points for the formation of helical gold nanowire.;
(i) Mechanism of atom insert on surface,
which introduces helical domain and domain boundary with point defects. 
(ii) The simple energy scaling theory that explains the net energy gain. 
The above two points are universal and not dependent on the details of simulations. 
In general, the inserted atom can be supplied not only from the inner shell 
but also from the outer area. \cite{IGUCHI2007}
We speculate that another candidate for the source of the atom supply is the wire ends,
if they are connected to electrode parts.

\section{Summary and future aspect}
\label{SEC-SUMMARY}

In summary, 
domain boundary formation is explored
for helical multishell gold nanowire.
The simulations were carried out for nanowires longer than 10 nm 
by quantum mechanical molecular dynamics simulation.
As results, 
the shear-like deformation on the (001)-type surface area 
introduces helical domains \cite{IGUCHI2007} and 
a domain boundary appears, 
typically with the supply of a defective atom pair from the inner shell, 
between two domains with the opposite shear directions.
The mechanism is explained quantitatively
by the analysis of local electronic structure and 
a general discussion is given 
by a simple energy scaling theory.

As a future aspect,
simulations should be carried out with larger samples 
that contain electrode parts, as pointed above.
We note that the simulation with realistic electrodes is important 
also for the transport property, 
particularly among helical multishell nanowires 
and other nanowires thicker than the monoatomic chain,
because of the possible interference effect  
at the connection part with electrodes. 
\cite{OHSHIMA2003, ONO2005, OSHIMA2006, KURUI2008, SHINAOKA}

A promising theoretical approach for larger quantum-mechanical 
simulations is  \lq order-$N$' method,
in which the computational time is \lq order-$N$' 
or proportional to the system size $N$. 
See articles cited in Ref.\cite{HOSHI2006} 
or a recent journal volume
that includes Ref.~\cite{FUJIWARA2008} 
and focuses on the order-$N$ methods.


\end{document}